\documentclass[11pts, prb,aps,twocolumn,showpacs,superscriptaddress]{revtex4-2} 
\usepackage{amsmath,amssymb,amsthm}
\usepackage{graphicx}
\usepackage{subfigure}
\usepackage{amsmath}
\usepackage{amsfonts,amssymb}
\usepackage{bm}
\usepackage{float}
\usepackage{color}
\usepackage{sidecap}
\allowdisplaybreaks
\usepackage{soul}
\setstcolor{red}
\usepackage[normalem]{ulem}
\usepackage{hyperref}
\hypersetup{colorlinks=true,breaklinks,urlcolor=blue,linkcolor=blue,citecolor=blue}

\begin{document}

\title{Dissolution of the non-Hermitian skin effect in one-dimensional lattices with linearly varying nonreciprocal hopping}

\author{Bo Hou}
\affiliation{Department of Physics, Capital Normal University, Beijing 100048, China}

\author{Han Xiao}
\affiliation{Department of Physics, Capital Normal University, Beijing 100048, China}

\author{Rong L\"u}
\affiliation{State Key Laboratory of Low-Dimensional Quantum Physics, Department of Physics, Tsinghua University, Beijing 100084, China}
\affiliation{Collaborative Innovation Center of Quantum Matter, Beijing 100084, China}

\author{Qi-Bo Zeng}
\email{zengqibo@cnu.edu.cn}
\affiliation{Department of Physics, Capital Normal University, Beijing 100048, China}

\begin{abstract}
We study the one-dimensional non-Hermitian lattices with linearly varying nonreciprocal hopping, where the non-Hermitian skin effect (NHSE) is found to be dissolved gradually as the strength of nonreciprocity increases. The energy spectrum under the open boundary condition is composed of real and imaginary eigenenergies when the nonreciprocal hopping is weak. Interestingly, the real eigenenergies form an equally spaced ladder, and the corresponding eigenstates are localized at the boundary with a Gaussian distribution due to NHSE. By increasing the nonreciprocity, the number of real eigenenergies will decrease while more and more eigenenergies become imaginary. Accompanied by the real-imaginary transition in the spectrum, the eigenstates are shifted from the boundary into the bulk of the lattice. When the nonreciprocity gets strong enough, the whole spectrum will be imaginary and the NHSE disappears completely in the system; i.e., all the eigenstates become Gaussian bound states localized inside the bulk. Our work unveils the exotic properties of non-Hermitian systems with spatially varying nonreciprocal hopping.
\end{abstract}
\maketitle
\date{today}

\section{Introduction}
During the past few years, the research on non-Hermitian physics has undergone rapid development~\cite{Cao2015RMP,Konotop2016RMP,Ganainy2018NatPhy,Ashida2020AiP,Bergholtz2021RMP}. Non-Hermitian Hamiltonians have been exploited to study a wide range of classical~\cite{Makris2008PRL,Klaiman2008PRL,Guo2009PRL,Ruter2010NatPhys,Lin2011PRL,Regensburger2012Nat,Feng2013NatMat,Peng2014NatPhys,Wiersig2014PRL,Hodaei2017Nat,Chen2017Nat} and quantum open systems~\cite{Brody2012PRL,Lee2014PRX,Li2019NatCom,Kawabata2017PRL,Hamazaki2019PRL,Xiao2019PRL,Wu2019Science,Yamamoto2019PRL,Yamamoto2019PRL,Naghiloo2019NatPhys,Matsumoto2020PRL}, and have unveiled many exotic phenomena that do not exist in traditional Hermitian systems. Since the operators are non-Hermitian, the eigenvalues, such as the eigenenergies of non-Hermitian Hamiltonians, are commonly complex. However, for the Hamiltonians that are $\mathcal{PT}$-symmetric~\cite{Bender1998PRL,Bender2002PRL,Bender2007RPP} or pseudo-Hermitian~\cite{Mostafazadeh2002JMP,Mostafazadeh2010IJMMP,Moiseyev2011Book,Zeng2020PRB1,Kawabata2020PRR,Zeng2021NJP}, the energy spectra can still be real. 

One of the most exotic phenomena in non-Hermitian systems is the accumulation of eigenstates at the system's boundaries, which is called the non-Hermitian skin effect (NHSE)~\cite{Yao2018PRL1,Yao2018PRL2}. The presence of NHSE results in a variety of phenomena that are absent in the corresponding Hermitian systems~\cite{Alvarez2018PRB,Alvarez2018EPJ,Lee2019PRB,Zhou2019PRB,Kawabata2019PRX,Song2019PRL,Okuma2020PRB,Xiao2020NatPhys,Yoshida2020PRR,Longhi2019PRR,Yi2020PRL,Claes2021PRB,Haga2021PRL,Zeng2022PRA,Zeng2022PRB}. For example, the band topology can be modified in a significant way, and the conventional principle of bulk-boundary correspondence in the Hermitian topological phase breaks down in non-Hermitian systems due to the NHSE~\cite{Yao2018PRL1,Yao2018PRL2,Kunst2018PRL,Jin2019PRB,Yokomizo2019PRL,Herviou2019PRA,Zeng2020PRB,Borgnia2020PRL,Yang2020PRL2,Zirnstein2021PRL,Zhang2023SciBull}. As a matter of fact, the emergence of NHSE itself also has a topological origin, which is closely connected to the point gap in the spectrum under the periodic boundary condition (PBC)~\cite{Okuma2020PRL,Zhang2020PRL}. The spectra of such systems are sensitive to the change of boundary conditions~\cite{Xiong2018JPC}, which inspires the designing of new quantum sensors~\cite{Budich2020PRL,Koch2022PRR}. In addition, the NHSE also influences the phenomenon of Anderson localization significantly~\cite{Hatano1996PRL,Shnerb1998PRL,Gong2018PRX,Jiang2019PRB,Zeng2020PRR,Liu2021PRB1,Liu2021PRB2}, where the spectra for extended and localized states exhibit different topological structures. So far, most studies mainly focus on the NHSE induced by homogeneous nonreciprocity in the hopping amplitude of the non-Hermitian lattice model, where the eigenstates are localized exponentially at the boundaries. If the nonreciprocity becomes spatially dependent, what will happen to the eigenenergies, eigenstates, and the NHSE remains unexplored. 

To answer these questions, in this paper, we study the one-dimensional (1D) lattices with linearly varying nonreciprocal hopping in the nearest-neighboring sites. When the nonreciprocity is weak, the energy spectrum is composed of real and imaginary eigenenergies. The real eigenenergies are found to be equally spaced with the eigenstates localized at one end of the 1D lattice due to the NHSE under the open boundary condition (OBC). Interestingly, differently from the systems with constant nonreciprocity, where the eigenstates are exponentially localized, here the eigenstates show Gaussian distributions at the boundary. The eigenstates corresponding to the imaginary eigenenergies are also Gaussian. However, they are not localized at the boundary of the lattice but become tightly bound states inside the bulk. As the strength of nonreciprocity increases, the real spectrum will disappear gradually, and all the eigenenergies become imaginary in the end. Accompanied by the real-imaginary transition in the OBC energy spectrum, the NHSE is dissolved completely, as all the eigenstates are shifted from the boundary into the bulk and become Gaussian bound states. Our work reveals the peculiar behaviors of energy spectra, eigenstates, and NHSE in the non-Hermitian 1D lattices with linearly varying nonreciprocal hopping. 

The rest of the paper is organized as follows. In Sec.~\ref{Sec2}, we will first introduce the model Hamiltonian of the 1D lattices with linearly increasing nonreciprocal hopping. In Sec.~\ref{Sec3}, we discuss the properties of the eigenenergy spectrum of the system. Then we will further explore the behaviors of the eigenstates and the dissolution of NHSE in Sec.~\ref{Sec4}. The last section (Sec.~\ref{Sec5}) is dedicated to a summary.

\section{Model Hamiltonian}\label{Sec2}
We introduce the 1D nonreciprocal lattice described by the following Hamiltonian
\begin{equation}\label{H}
  \begin{aligned}
	H &= \sum_{j=1}^{L-1} t_j c_{j}^\dagger c_{j+1} +t_j^\prime c_{j+1}^\dagger c_{j} \\ 
	&= \sum_{j=1}^{L-1} \left( t +  \gamma j \right ) c_{j}^\dagger c_{j+1} + \left( t - \gamma j \right) c_{j+1}^\dagger c_{j}.
  \end{aligned}
\end{equation}
Here $c_j$ ($c_j^\dagger$) is the annihilation (creation) operator of spinless fermions at the $j$th site. $t_j=\left( t+\gamma j \right)$ and $t_j^\prime=\left( t-\gamma j \right)$ are the backward and forward hopping between the nearest-neighboring sites, which vary linearly along the system. $t$ is the constant hopping amplitude and is set to be $1$ as the energy unit throughout this paper. $\gamma$ indicates the strength of nonreciprocity in the hopping terms. Both $\gamma$ and $t$ are real numbers. $L$ is the number of lattice sites.  Distinguished from the models in previous studies, where the nonreciprocal hopping is homogeneous along the whole lattice, the nonreciprocal hopping here is site-dependent and increases linearly in the system. In the following sections, we will check how the linearly increasing nonreciprocity will affect the properties of the energy spectrum and NHSE in 1D lattices.

\section{Eigenenergy spectrum}\label{Sec3}
\begin{figure}[t]
	\includegraphics[width=3.3in]{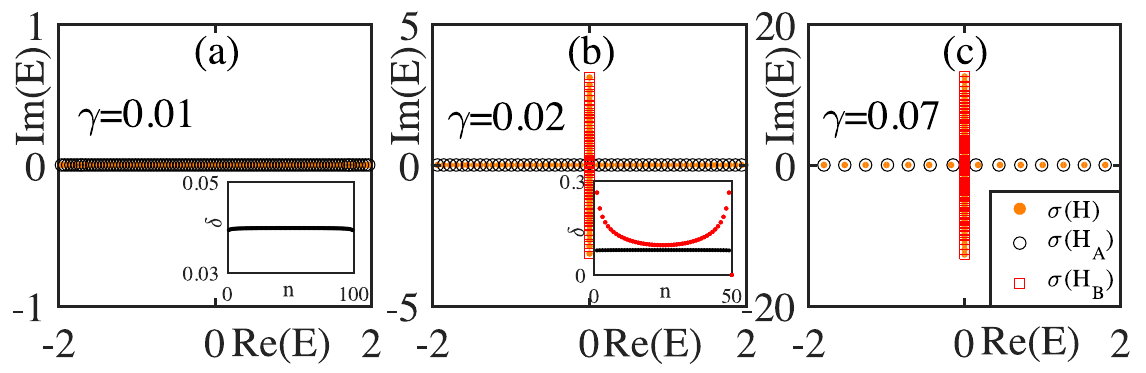}
	\caption{(Color online) The eigenenergy spectrum under OBC of the 1D lattice with different linearly increasing nonreciprocity: (a) $\gamma=0.01$, (b) $\gamma=0.02$, and (c) $\gamma=0.07$. The black circles and red squares represent the energy spectrum of $h_A$ and $h_B$ in the matrix $h$ in Eqs.~(\ref{h2}) and (\ref{h3}). The insets in (a) and (b) show the level spacings of the real (black dots) and imaginary (red dots) eigenenergies, respectively. Here the lattice size is $L=100$.}
	\label{fig1}
\end{figure}

We first check the energy spectrum of the system. For a 1D lattice described by Eq.~(\ref{H}) with $L$ sites under OBC, the model Hamiltonian can be represented by a $L\times L$ tridiagonal matrix. Since the Hamiltonian is non-Hermitian, the eigenvalues of the matrix are normally complex. Interestingly, for the Hamiltonian shown in Eq.~(\ref{H}), we find that the eigenenergy spectrum is composed of real and imaginary energies. To illustrate this, we can make a similarity transformation to the matrix by $h=D^{-1}HD$ with $D$ being a diagonal matrix: $D=diag(d_1,d_2,\cdots,d_L)$, where the diagonal elements for the case with $|t/\gamma| \geq L$ are given by 
\begin{equation}\label{dj}
	d_j = \left\{
	\begin{array}{c}
		1, \qquad j = 1 \\
		\sqrt{\frac{t^\prime_{j-1} t^\prime_{j-2} \cdots t^\prime_1}{t_{j-1}t_{j-2} \cdots t_1}}, \qquad j = 2,3,\cdots,L
	\end{array}
	\right.
\end{equation}
Then the non-Hermitian Hamiltonian matrix $H_1$ with $|t/\gamma| \geq L$ is transformed into the following Hermitian matrix
\begin{widetext}\label{H_1}
	\begin{equation}\label{h1}
		h_1 = D^{-1}H_1 D = \left(
		\begin{array}{ccccc}
			V_1 & sgn(t_1)\sqrt{t_1 t^\prime_1} & 0 & \cdots & 0 \\
			sgn(t_1)\sqrt{t_1 t^\prime_1} & V_2 & sgn(t_2)\sqrt{t_2 t^\prime_2} & \cdots & 0 \\
			0 & sgn(t_2)\sqrt{t_2 t^\prime_2} & V_3 & sgn(t_3)\sqrt{t_3 t^\prime_3} & \cdots \\
			\vdots & \vdots & \vdots & \ddots & \vdots \\
			0 & 0 & 0 & \cdots & V_L \\
		\end{array}
		\right).
	\end{equation}
\end{widetext} 
Thus, the spectrum of the Hamiltonian $H_1$ is purely real. In Fig.~\ref{fig1}(a), we present the eigenenergies of a lattice under OBC with $\gamma=0.01$ and $L=100$, which are all real as expected. 

However, when $|t/\gamma|<L$, the situation becomes more complicated. For instance, if we have $\gamma>0$ and $|t/\gamma|=m<L$ with $m$ being a positive integer, then the backward and forward hopping between the $m$th and $(m+1)$th site will be $t_m = t+mj$ and $t_m^\prime=0$, respectively. The Hamiltonian can be represented as a block matrix as follows
\begin{equation}\label{H2}
	H_2 = \left(
	\begin{array}{c|c}
		H_A & t+mj \\
		\hline
		0 & H_B
	\end{array}
	\right),
\end{equation}
where $H_A$ is an $m\times m$ dimensional matrix and $H_B$ is an $(L-m)\times (L-m)$ dimensional matrix. Then the eigenenergies of $H$ are determined by $H_A$ and $H_B$. By performing the similarity transformation using the $D$ matrix with the following diagonal elements
\begin{equation}\label{dj2}
		d_j = \left\{
	\begin{array}{c}
		1, \qquad j = 1 \\
		\sqrt{\frac{t^\prime_{j-1} t^\prime_{j-2} \cdots t^\prime_1}{t_{j-1}t_{j-2} \cdots t_1}}, \qquad j = 2,3,\cdots,m \\
		1, \qquad j=m+1 \\
		\sqrt{\frac{t^\prime_{j-1} t^\prime_{j-2} \cdots t^\prime_{(m+1)}}{t_{j-1}t_{j-2} \cdots t_{(m+1)}}}, \qquad j = m+2,m+3,\cdots,L
	\end{array}
	\right.,
\end{equation} 
we can transform $H_A$ into a Hermitian matrix and $H_B$ into a anti-Hermitian one. The Hamiltonian matrix becomes
\begin{equation}\label{h2}
	h_2 = D^{-1} H_2 D = \left(
		\begin{array}{c|c}
		h_A & c \\
		\hline
		0 & i h_B
	\end{array}
	\right).
\end{equation}
Here, both $h_A$ and $h_B$ are Hermitian matrices, and $c$ is a nonzero real number. The eigenenergies of $H$ are given by the eigenvalues of $h_A$ and $i h_B$, which means that there are $m$ real eigenenergies and $(L-m)$ imaginary eigenenergies. Figure ~\ref{fig1}(b) shows the spectrum $\sigma(H)$ of the original Hamiltonian matrix in Eq.~(\ref{H}) and those of $h_A$ and $i h_B$ [i.e., $\sigma(h_A)$ and $\sigma(h_B)$] when $\gamma=0.02$. We can see that they are perfectly matched with each other. Notice that if $t/\gamma >L$, then we have $h_A = h_1$ and thus the real spectrum of $h_A$ still matches the one of $H$, as indicated by the black circles and yellow dots in Fig.~\ref{fig1}(a). From the above analysis, we find that in the case with $t/\gamma$ being an integer, the spectrum of the 1D lattice can be divided into two independent parts, where $H_A$ (or $h_A$) represents the part with $t_j^\prime=(t-\gamma j) <0$ while $H_B$ (or $h_B$) represents the part with negative $(t-\gamma j) >0$. However, if $t/\gamma$ is not an integer, the model Hamiltonian is  
\begin{equation}
	H_3 = \left(
	\begin{array}{c|c}
		H_A & t+\gamma s \\
		\hline
		t-\gamma s & H_B
	\end{array}
	\right),
\end{equation}
with $s=\lfloor t/\gamma \rfloor$ being the largest integer when $(t-\gamma s)$ is positive. Then we can still make the similarity transformation using the diagonal matrix $D$ with elements shown in Eq.~(\ref{dj}) to transform the model Hamiltonian matrix into a block form as 
\begin{equation}\label{h3}
	h_3 = D^{-1} H_3 D = \left(
	\begin{array}{c|c}
		h_A & a \\
		\hline
		b & i h_B
	\end{array}
	\right),
\end{equation}
where $a$ and $b$ are two nonzero numbers. Differently from $H_2$, now the system cannot be divided into two independent parts. As we can see from Fig.~\ref{fig1}(c), the eigenenergy spectrum still composes of real and imaginary values,  but the eigenvalues of $H_A$ and $H_B$ cannot fully match the ones of $H_3$.

\begin{figure}[t]
	\includegraphics[width=3.3in]{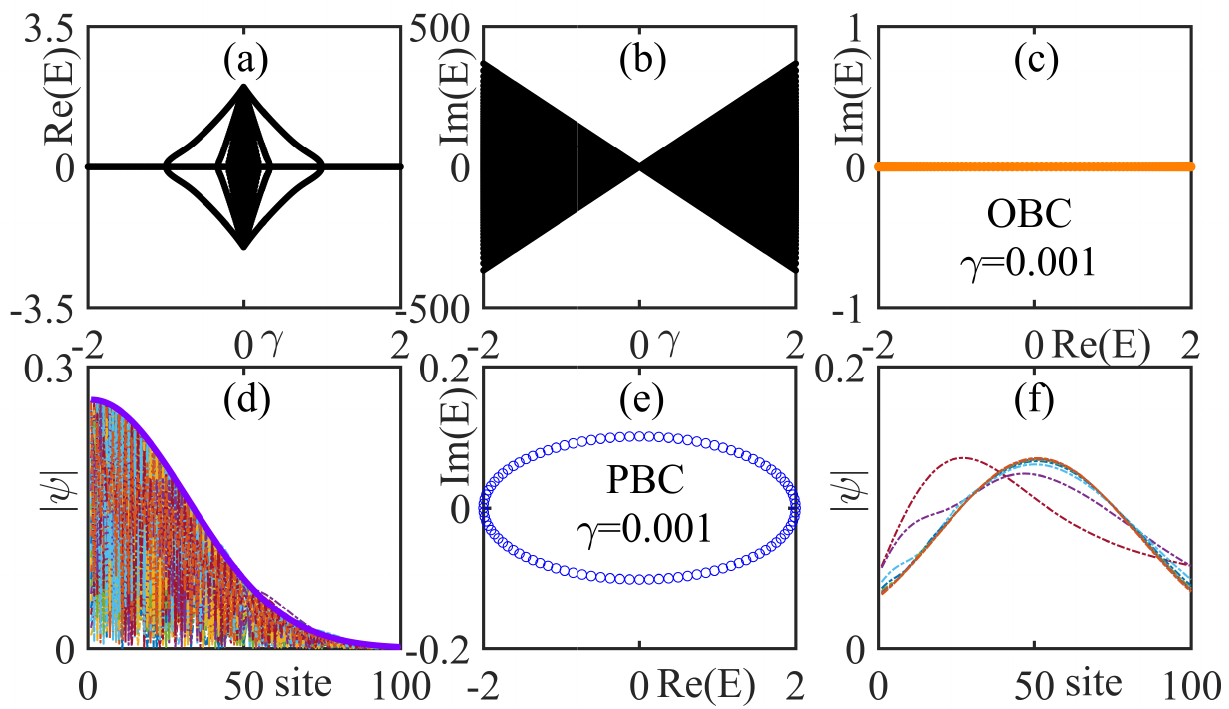}
	\caption{(Color online) (a) Real and (b) imaginary parts of the eigenenergies of the 1D lattice under OBC described by Eq.~(\ref{H}) as a function of $\gamma$. (c)-(f) show the energy spectra and the distribution of eigenstates under OBC and PBC when $\gamma=0.001$. The purple solid line in (d) represents the Gaussian envelope function. Here the lattice size is  $L=100$.}
	\label{fig2}
\end{figure}

In Figs.~\ref{fig2}(a) and \ref{fig2}(b), we plot the real and imaginary parts of the energy spectrum as a function of the nonreciprocity $\gamma$. It is clear that when $|\gamma|>t$, the spectrum becomes purely imaginary. The reason is that when the nonreciprocity $\gamma>t$ (or $\gamma<-t$), all the terms $(t-\gamma j)$ [or $(t+\gamma j)$] in the Hamiltonian are negative, then the matrix after similarity transformation is anti-Hermitian, and the spectrum will be imaginary. 

Another interesting feature in the energy spectrum of the 1D lattice with linearly increasing nonreciprocity is that the real eigenenergies are almost equally spaced. To see that, we sort the real eigenenergies in order from the smallest to the largest and get a set $\{ E_{nR} \}$. As to the imaginary eigenenergies, we sort them by their imaginary part and get another set $\{ E_{nI} \}$. Then the level spacing is defined as 
\begin{equation}
	\delta_{n,n+1} = E_{n+1} - E_n,
\end{equation}
where $E_n$ is the $n$th eigenenergy in $\{ E_{nR} \}$ or $\{ E_{nI} \}$. The insets in Figs.~\ref{fig1}(a) and \ref{fig1}(b) show the level spacings of the real and imaginary eigenenergies. We can see that the level spacings of the real eigenenergies are almost constant, indicating that the real spectrum forms an equally spaced ladder, similar to the Wannier-Stark ladder in the 1D lattices imposed by a uniform external field~\cite{Wannier1962RMP,Fukuyama1973PRB,Emin1987PRB}. On the other hand, the behavior of the imaginary eigenenergies is quite different, where the level spacings are not constant. So, the real and imaginary spectra behave differently in this model. It will be interesting to check whether the eigenstates corresponding to the real and imaginary energies will also exhibit distinctive behaviors.

The spectrum of our model also behaves differently from the non-Hermitian systems with constant nonreciprocity. For instance, in the 1D Hatano-Nelson model described by the Hamiltonian $H_{HN}=\sum_j (t+\gamma) c_{j+1}^\dagger c_j + (t-\gamma) c_j^\dagger c_{j+1}$, the eigenenergies under OBC are real when $|\gamma|<t$ but become imaginary when $|\gamma|>t$. The real-imaginary transition only depends on the strength of $\gamma$ and is independent of the system size. Similar behaviors can also be observed in other non-Hermitian lattices with constant asymmetric hopping such as the 1D Su-Schrieffer-Heeger model in Ref.~\cite{Yao2018PRL1} and the mosaic nonreciprocal lattices in Ref.~\cite{Zeng2022PRB}. However, for the model we studied here, since the nonreciprocal hopping increases linearly and thus depends on the lattice size, the real-imaginary transition spectrum also becomes size dependent. If both $\gamma$ and $L$ are small, the spectrum is purely real. When the system size gets larger than a critical number, there will be both real and imaginary eigenenergies. If $\gamma$ becomes very strong, then the spectrum is purely imaginary. So, the real-imaginary transition in the spectrum of our model is determined by nonreciprocity $\gamma$ and lattice size $L$. 

While we have mainly discussed the cases with $\gamma>0$ in the above, the method can also be used for the cases with $\gamma<0$, where similar conclusions will be obtained. So, in the 1D lattices with linearly increasing nonreciprocal hopping, we find that as the nonreciprocity gets strong or as $j$ increases, the eigenenergy will undergo a real-imaginary transition. In the next section, we will further investigate how the linearly increasing nonreciprocity will influence the behaviors of eigenstates and the NHSE.

\section{NHSE and tightly bound states}\label{Sec4}
As the hopping amplitudes between the nearest-neighboring sites are nonreciprocal in our model, we can expect the emergence of NHSE, where the eigenstates accumulate at the boundaries of the 1D lattice. As shown in Figs.~\ref{fig2} (c) and \ref{fig2}(d), the eigenenergies for lattices with $\gamma=0.001$ are real, and all the eigenstates are localized at the left boundary. However, differently from the previous models with constant nonreciprocity in the hopping terms, such as the famous Hatano-Nelson model, where the eigenstates are exponentially localized at the boundary, here we find that the eigenstates form a Gaussian distribution instead. More interestingly, the accumulation of eigenstates forms a Gaussian bell shape, as indicated by the thick solid purple line in Fig.~\ref{fig2}(d), which can be approximately described by a Gaussian envelope function as
\begin{equation}\label{Gaussian}
	\phi(x) = max(|\psi|) e^{-\frac{|\gamma|}{2} (x-x_0)^2},
\end{equation}
where $max(|\psi|)$ represents the largest component in the distribution of all eigenstates and $x_0$ is the center of the distribution, which is $1$ here. So, due to the spatially linearly varied feature in the nonreciprocal hopping, the eigenstates and the NHSE behaves differently from the systems with constant nonreciprocity. 

On the other hand, it is well known that the existence of NHSE under OBC is closely connected to the point gap in the PBC spectrum~\cite{Okuma2020PRL,Zhang2020PRL}. In Figs.~\ref{fig2}(e) and \ref{fig2}(f), we present the eigenenergies and eigenstates under PBC. The eigenstates now are extended over the whole lattice. The eigenenergies form a closed loop in the complex energy plane and can be characterized by a nonzero winding number. This is the topological origin of the NHSE under OBC. 

\begin{figure}[t]
	\includegraphics[width=3.3in]{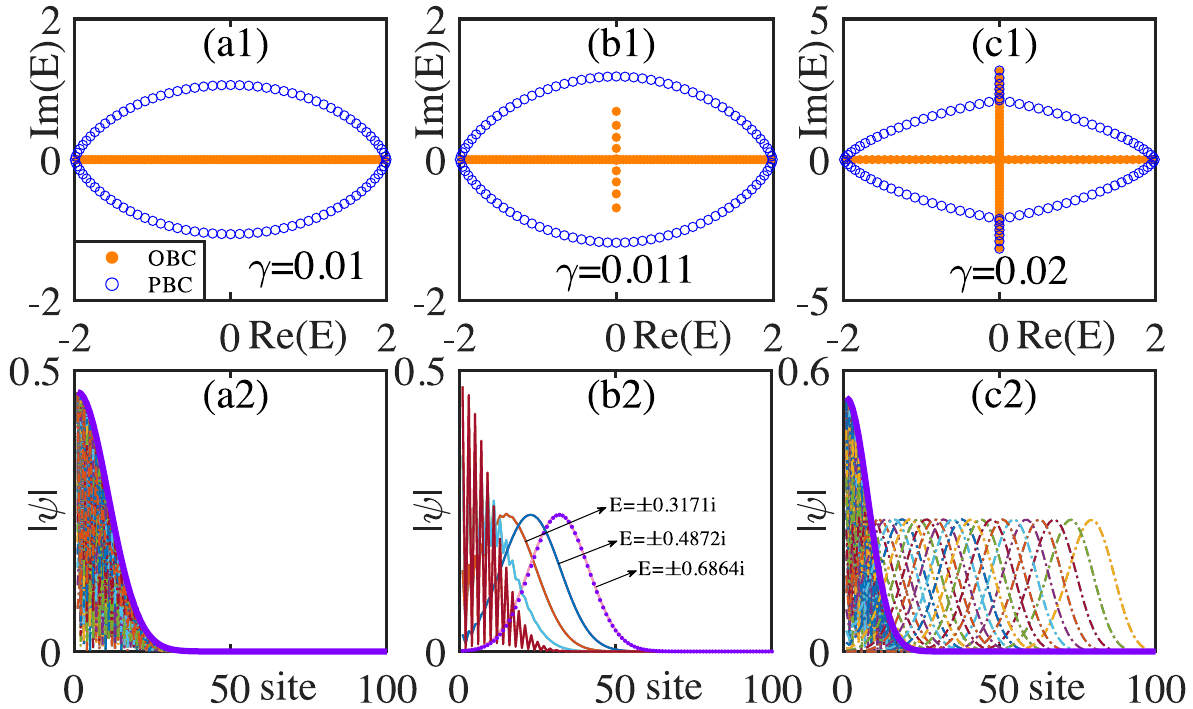}
	\caption{(Color online) (a1)-(c1) Energy spectrum under OBC (yellow solid dots) and PBC (blue circles) of the 1D lattice with different linearly increasing nonreciprocal hopping. As the strength of $\gamma$ increases, the OBC spectrum undergoes a real-imaginary transition. The distributions of the corresponding eigenstates under OBC are given in (a2)-(c2). The eigenstates under OBC are shifted from the boundary into the bulk and become tightly bound states as the real eigenenergies change to imaginary ones. The purple dots represent the values of the Gaussian function at $j$th sites, which are well matched with the eigenstates with eigenenergies $\pm 0.6864i$. Other parameters: $L=100$.}
	\label{fig3}
\end{figure}

As stated in the above section, when the nonreciprocity increases, the energy spectrum of the system under OBC will undergo a real-imaginary transition. For the lattice with size $L=100$, the hopping terms $(t-\gamma j) \geq 0$ when $0<\gamma <0.01$, and the OBC spectrum will be purely real, as shown in Fig.~\ref{fig3}(a1) for the case with $\gamma=0.01$. The eigenstates accumulate at the left end of the lattice and the whole distribution can be encapsulated by a Gaussian envelope function as represented by the solid purple line. If we increase the strength of nonreciprocity a little, for instance, we set $\gamma = 0.011$, then $10$ out of the $100$ eigenenergies will become imaginary, as shown in Fig.~\ref{fig3}(b1) (notice that there are two imaginary energies that are very close to zero). Now, the PBC spectrum again forms a closed loop, implying that there will be NHSE under OBC. It seems that all eigenstates will still localize at the left end. However, we find that the eigenstates corresponding to the imaginary eigenenergies under OBC will not be localized at the boundary. Instead, they are shifted into the bulk, as shown in Fig.~\ref{fig3}(b2). The larger the imaginary part is, the further will the corresponding eigenstate be moved away from the boundary. Moreover, these eigenstates are still Gaussian. For instance, for the two eigenstates $v_{1,2}$ with eigenenergies $\pm 0.6864i$, the distribution of the wave functions can be well approximated by using Eq.~(\ref{Gaussian}), where $max(|\psi|)$ now is the largest component of $|v_1|$ or $|v_2|$ and $x_0=32$. Thus the states with imaginary energies become tightly bound states in the bulk, and the NHSE is partially dissolved. Even though the PBC eigenenergies enclose the whole OBC spectrum and the point gap still exists, not all the eigenstates are accumulated at the boundary under OBC. 

If we increase the strength of the nonreciprocity in the hopping terms further, then more and more eigenenergies under OBC will become imaginary. Correspondingly, more and more eigenstates will be shifted into the bulk as a tightly bound state, as shown in Fig.~\ref{fig3}(c). Notice that in this case, some of the eigenenergies under PBC also become imaginary. When $\gamma$ becomes strong enough, all the eigenenergies become imaginary, and the PBC spectrum will be identical to the OBC spectrum. The closed loop formed by the PBC eigenenergies disappears, and there is no point gap anymore. Accordingly, no states will be localized at the end of the 1D lattice under OBC; i.e., the NHSE totally disappears. The disappearance of NHSE can be qualitatively understood as follows. When $\gamma$ is very strong, the constant term $t$ in the forward and backward hopping can be ignored. Then the hopping amplitudes are almost the same for the two directions, leading to the dissolution of NHSE. So, by tuning the strength of the linearly increasing nonreciprocity, the NHSE can be dissolved gradually, which is accompanied by the real-imaginary transition in the energy spectrum under OBC. 

\begin{figure}[t]
	\includegraphics[width=3.0in]{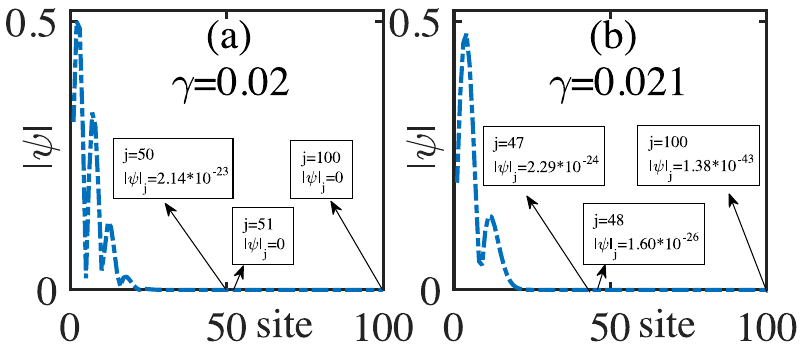}
	\caption{(Color online) The distribution of the eigenstates with real eigenenergies of the 1D nonreciprocal lattice when (a) $\gamma=0.02$ and (b) $\gamma=0.021$. The lattice size here is $L=100$.}
	\label{fig4}
\end{figure}

The behaviors of the eigenstates of our model described above are totally different from the 1D lattice with constant nonreciprocity, where the eigenstates are localized at the boundaries exponentially under OBC. The situation also differs from the system with disorders, where localized eigenstates arise in bulk due to the Anderson localization phase transition. The model studied here is disorder-free. The emergence of tightly bound states is similar to the Wannier-Stark (WS) localization in 1D lattices with a uniform external field~\cite{Wannier1962RMP,Fukuyama1973PRB,Emin1987PRB}, which results in a linear variation in the onsite potential in the model Hamiltonian. It is known that in the WS localization phenomenon, the eigenenergies will form an equally spaced ladder, and the eigenstates are tightly bound states. However, our model differs from the WS Hamiltonian in that the linear variation is only added in the hopping terms instead of the onsite terms. Moreover, the bound states in the WS localization are localized inside the bulk, which is not the case for our model since the linearly increasing hopping is asymmetric and the bound states can be localized at the boundary by NHSE . Here the more peculiar feature is that the equally spaced ladder only exists in the real spectrum with the corresponding eigenstates localized at the boundary due to NHSE, while the tightly bound states in the bulk have imaginary eigenenergies. In this sense, the ladder in the spectrum and the tightly bound states in the bulk are separated in our model. 

On the other hand, by taking a more careful investigation of the eigenstates with real eigenenergies, we can find that the behaviors of these states localized at the boundary depend on whether $|t/\gamma|$ is an integer or not. If we have $|t/\gamma|=m<L$, the Hamiltonian matrix takes the form of Eq.~(\ref{H2}), where the eigenenergies are determined by $H_A$ and $H_B$. Let us further show that the eigenstates for the real eigenenergies are also only determined by $H_A$. In Fig.~\ref{fig4}(a), we present the distribution of one of the eigenstates with real energy for $H_2$ with $\gamma=0.02$ and $L=100$. We can see that for $j \geq 51$, the $j$th component of the eigenstate becomes zero, meaning that the eigenstates with real eigenenergies in this case are fully restricted in the region with $j \leq m$. These states are also the eigenstates of $H_A$. To see this, suppose that $\psi_A$ is the eigenstate of $H_A$ with eigenenergy $E_A$; then we can construct a state vector as $\psi = (\psi_A \quad \boldmath{0})^T$ and we can prove that $H_2 \psi = E_A \psi$. Thus, the eigenenergies and eigenstates are the same for $H_A$ and $H_2$, and they are not dependent on the rest of the lattice at all. However, if $t/\gamma$ is not an integer, then such decomposition will not hold. In Fig.~\ref{fig4}(b), we also present the profile of one eigenstate with real energy from the system with $\gamma=0.021$ and $L=100$. We can find that components of the state at the lattice sites with $j> \lfloor t/\gamma \rfloor$ are very small but not zeros; thus they cannot be taken as an independent part.

\begin{figure}[t]
	\includegraphics[width=3.0in]{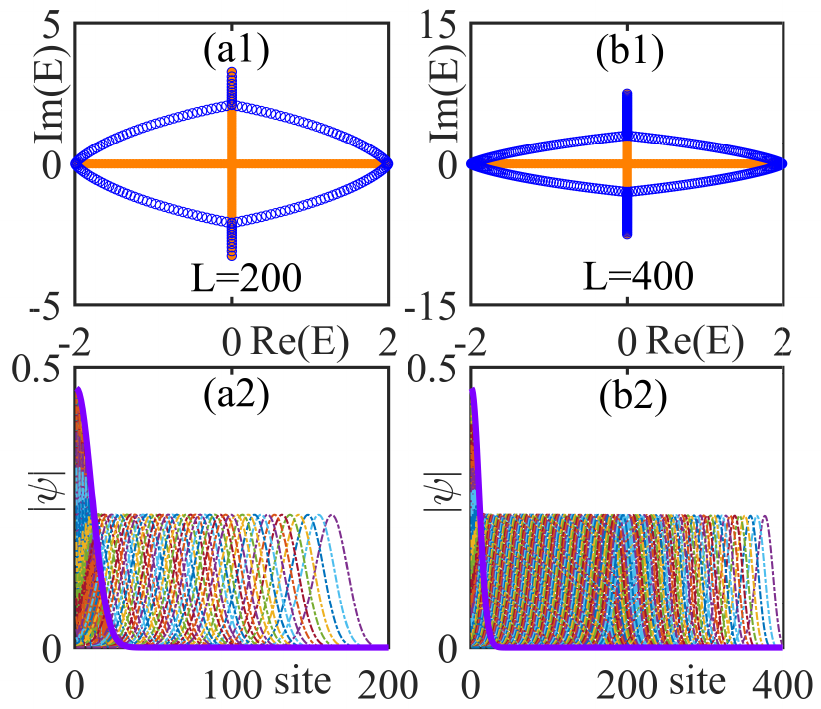}
	\caption{(Color online) Energy spectra of the 1D lattice with $\gamma=0.01$ and different sizes: (a1) $L=200$, (b1) $L=400$. The yellow solid dots and blue circles represent the OBC and PBC spectra, respectively. (a2) and (b2) show the corresponding eigenstates under OBC.}
	\label{fig5}
\end{figure}

Since the nonreciprocal hopping increases linearly with $j$, there will always be negative hopping terms in the model Hamiltonian, as long as the lattice is long enough. Thus in the thermodynamic limit, there will always be imaginary eigenenergies in the spectrum with the corresponding eigenstates shifted from the boundary into the bulk. In Fig.~\ref{fig5}, we present the energy spectrum and the profile of eigenstates for the 1D lattice with $L=200$ and $400$, respectively. Here, the nonreciprocity $\gamma=0.01$. The hopping between the first 100 sites will be positive, while the rest of the forward hopping terms are negative. Then some of the eigenstates will be localized at the boundary due to NHSE, and the others are tightly bound states in the bulk. Moreover, according to the above analysis, as $t/\gamma=100$ is an integer, the real eigenenergies and the corresponding eigenstates are only determined by the first $100$ sites.

\section{Summary}\label{Sec5}
In this paper, we study the 1D non-Hermitian lattices with linearly increasing nonreciprocal hopping between the nearest-neighboring sites. We find that due to the spatially varying nonreciprocity, the eigenenergies and eigenstates behave quite differently from those of non-Hermitian systems with constant nonreciprocity. When the nonreciprocity is weak, some of the eigenenergies under OBC remain real and form an equally spaced ladder. The corresponding eigenstates are localized at one end of the 1D lattice due to the NHSE and exhibit a Gaussian distribution. The rest of the eigenenergies are imaginary with the eigenstates shifted from the boundary into the bulk and forming tightly bound states. As the strength of nonreciprocity increases, the energy spectrum undergoes a real-imaginary transition, accompanied by the shifting of eigenstates from the boundary to the bulk. Thus, the NHSE is dissolved gradually. When the nonreciprocity becomes strong enough, there will be no NHSE in the system, and all the states are Gaussian bound states in the bulk. It is quite interesting to see that the increment in nonreciprocal hopping does not necessarily lead to the enhancement of NHSE. On the contrary, the model we studied here shows that the linear variation in the nonreciprocal hopping terms can dissolve the NHSE. Our work reveals the exotic properties of non-Hermitian lattices with spatially varying nonreciprocity and opens a door for future studies on such systems. As to the experimental realization, it has been reported that nonreciprocal hopping can be realized by using photonic coupled resonant optical waveguides, where judicious optical gain and loss elements in the coupling link rings can be designed~\cite{Zhu2020PRR}. The variation in the hopping terms can be realized by tuning the distance between the neighboring waveguides.

\begin{acknowledgments}
This work is supported by the NSFC (Grant No. 12204326), Beijing Natural Science Foundation (Grant No. 1232030), and R\&D Program of the Beijing Municipal Education Commission (Grant No. KM202210028017).  R.-L. is supported by the NSFC under Grant No. 11874234, the National Key Research and Development Program of China
(Grant No. 2018YFA0306504), and the Ministry of Science and Technology of China (Grand No. 2023YFC2205800).
\end{acknowledgments}

\end{document}